\documentclass[aps,prd,amsfonts,superscriptaddress,nofootinbib,
               twocolumn,byrevtex,tightenlines,showpacs]{revtex4}
\usepackage{bm,hyperref,mathrsfs}
\newcommand{\getp}{{\get^\parallel}}
\newcommand{\getn}{{\get^\perp}}

\newcommand{\vc}[1]{{\bm{#1}}}
\newcommand{\der}{\partial}

\DeclareMathSymbol{\mg}{\mathrel}{symbols}{"1D} 

\newcommand{\mx}[1]{\bm{#1}}

\newcommand{\Bnabla}{\bm{\nabla}}

\newcommand{\half}{\frac{1}{2}}
\newcommand{\Id}{\openone}

\newcommand{\ra}{\rightarrow}
\newcommand{\inv}{^{-1}}
\newcommand{\lh}{\left(}
\newcommand{\rh}{\right)}
\newcommand{\labl}[1]{\label{#1}}
\newcommand{\eqref}[1]{(\ref{#1})}

\renewenvironment{matrix}{%
  \hskip -\arraycolsep\array{cccccccccc}%
}{%
  \endarray \hskip -\arraycolsep
}
\newcommand{\beq}{\begin{equation}}
\newcommand{\eeq}{\end{equation}}
\newcommand{\bea}{\begin{eqnarray}}
\newcommand{\eea}{\end{eqnarray}}
\renewenvironment{pmatrix}{\left(\matrix}{\endmatrix\right)}
\newcommand{\equ}[1]{\begin{eqnarray} #1 \end{eqnarray}}

\newcommand{\ga}{\alpha}
\newcommand{\gb}{\beta}
\newcommand{\ggam}{\gamma}
\newcommand{\gd}{\delta}
\renewcommand{\ge}{\epsilon}

\newcommand{\gz}{\zeta}
\newcommand{\get}{\eta}
\newcommand{\gth}{\theta}

\newcommand{\gk}{\kappa}
\newcommand{\gl}{\lambda}
\newcommand{\gm}{\mu}
\newcommand{\gn}{\nu}
\newcommand{\gx}{\xi}
\newcommand{\gp}{\pi}

\newcommand{\gs}{\sigma}
\newcommand{\gt}{\tau}

\newcommand{\gf}{\phi}
\newcommand{\gvf}{\varphi}
\newcommand{\gc}{\chi}

\newcommand{\gG}{\Gamma}
\newcommand{\gD}{\Delta}

\newcommand{\gL}{\Lambda}

\newcommand{\gP}{\Pi}

\newcommand{\gF}{\Phi}
\newcommand{\gPs}{\Psi}
\newcommand{\gO}{\Omega}
\newcommand{\cA}{{\mathcal A}}
\newcommand{\cB}{{\mathcal B}}

\newcommand{\cD}{{\mathcal D}}

\newcommand{\cH}{{\mathcal H}}

\newcommand{\cM}{{\mathcal M}}

\newcommand{\cQ}{{\mathcal Q}}

\newcommand{\cU}{{\mathcal U}}

\newcommand{\tp}{{\tilde p}}

\newcommand{\tv}{{\tilde v}}

\newcommand{\tx}{{\tilde x}}

\newcommand{\tB}{{\tilde B}}
\newcommand{\tC}{{\tilde C}}

\newcommand{\tga}{{\tilde\alpha}}
\newcommand{\tgb}{{\tilde\beta}}
\newcommand{\tgg}{{\tilde\gamma}}

\newcommand{\tgk}{{\tilde\kappa}}

\newcommand{\tgm}{{\tilde\mu}}

\newcommand{\tgp}{{\tilde\pi}}

\newcommand{\tgvf}{{\tilde\varphi}}
\newcommand{\tgc}{{\tilde\chi}}

\newcommand{\tgPs}{{\tilde\Psi}}

\newcommand{\Bgd}{\bm{\delta}}

\newcommand{\Bgf}{\bm{\phi}}

\newcommand{\Bget}{\bm{\eta}}

\newcommand{\BgO}{\bm{\Omega}}

\newcommand{\Bgfc}{\Bgf_\chi}
\newcommand{\gvfc}{\gvf_\chi}
\newcommand{\gac}{\ga_\chi}
\newcommand{\gksq}{\kappa_0^2}
\newcommand{\cDt}{\cD_t}
\newcommand{\cDe}{\cD_\eta}
\newcommand{\Bq}{\bm{q}}
\newcommand{\Bk}{\bm{k}}

\newcommand{\geH}{\ge_\cH}
\newcommand{\getH}{\get_\cH}
\newcommand{\gthH}{\gth_\cH}
\newcommand{\zH}{z_\cH}
\begin{document}

\title{Cosmological perturbations in multiple-field inflation}

\author{Joydev Lahiri}
\email{joy@veccal.ernet.in}
\affiliation{Variable Energy Cyclotron Centre, 1/AF Bidhan Nagar
Kolkata 700 064, India}
\author{Gautam Bhattacharya}
\email{gautam@theory.saha.ernet.in}
\affiliation{Saha Institute of Nuclear Physics, 1/AF Bidhan Nagar
Kolkata 700 064, India}

\date{\today}

\begin{abstract}
\noindent We analyze cosmological perturbations to the linear
order in the context of inflation with an
arbitrary number of scalar fields. The fields take values on a
non-trivial manifold with a positive definite metric
and are non-minimally coupled to Einstein gravity. The 
perturbations are decomposed into three different types. The
scalar-type perturbations are presented in a gauge-ready form without
fixing the temporal gauge condition, as well as in terms of 
gauge-invariant variables. The gauge-ready method enables us to
impose different gauge conditions which are most suitable to the
problem at hand. We quantize the scalar perturbations
and obtain the solutions.
\end{abstract}


\maketitle

\section{Introduction}
\labl{intro}
It is well known that the cosmological inflation, an epoch of accelarated
expansion, provides a causal mechanism for the generation and evolution of
large-scale structure formation in the Universe \cite{Guth,Lindebook,
Liddlebook}.
Inflation explains a number of puzzles of the Big-Bang theory, such as
homogeneity, the isotropy of the Cosmic Microwave Background Radiation (CMBR)
and the flatness of space-like sections.  As an added bonus, it connects
cosmology with high-energy physics, thus forming a cosmic laboratory 
where one may probe physics beyond the Standard Model. The very high
accuracy CMBR data recently obtained by the WMAP satellite \cite{WMAP} 
provide a new impetus to compare the predictions of cosmological
inflation with observations and hopefully to discover new physics
in the very high energy regime.

At the simplest level, the inflationary scenario is implemented
by assuming that the matter is described by a single scalar field,
the \textit{inflaton}, which is a special case of a perfect 
fluid \cite{Guth,Single}. As the early Universe undergoes inflation,
quantum fluctuations of the scalar fields are generated which become
classical after crossing the event horizon. During the decelaration phase
they re-enter the horizon and seed the observed density perturbations.
Cosmological perturbations in a single field
inflation has been thoroughly investigated in the past
\cite{Liddlebook,Mukhanovetal}. 

Despite initial successes, the single component inflation also has its
drawbacks. It was realized quite early that, in its original form, the
inflationary scenario suffers from what is called the \textit{graceful-exit}
problem \cite{Guth,Exit}. Linde \cite{Linde90} showed that in order to
achieve sufficient inflation consistent with the observed
density perturbations, before the Universe exits from the inflationary epoch,
one requires at least \textit{two} scalar fields without modifying
Einstein gravity, and without sacrificing natural initial conditions.

Of course, there are other motivations for incorporating multiple scalar
fields contributing to the dynamics of inflation. When constructing models
of inflation inspired by particle physics theories such as
low energy effective supergravity derived from superstrings, one obtains
many scalar fields (see \cite{Lythrep} for a recent review). Thus it is
necessary to have a general framework for handling cosmological perturbations
in a situation where the matter sector consists of an arbitrary number
of scalar fields. A method for treating density perturbations
in multicomponent inflation was proposed in \cite{Tent}, but see also
\cite{Stewart1,Stewart2,Stewart3}.

The study of cosmological perturbations was initiated by Lifshitz \cite{Lifs}
in 1946 when he analyzed hydrodynamical fluid perturbations in Einstein
gravity. He assumed a particular gauge which is now known as the synchronous
gauge. This gauge does not completely fix the gauge degrees of freedom and
the spurious gauge modes have to be properly sorted out in order to obtain
correct results. Later on, the zero-shear gauge was used by Harrison 
\cite{Harris} and the comoving gauge by Nariai \cite{Nariai}. However, it
was the seminal paper by Bardeen \cite{Bardeen80} which helped put
cosmological perturbations on a proper footing. He introduced a number of
gauge-invariant variables to the linear order, in terms of which the 
perturbations became much simpler to analyze. Reviews of cosmological
perturbations may be found in \cite{Mukhanovetal} and \cite{Kodamasasaki}. 

A different approach to cosmological perturbations was elaborated by
Hwang and colleagues in a 
series of papers \cite{Hwang1,Hwang2,Hwang3,Hwang4},
following a suggestion by Bardeen \cite{Bardeen88}, that
rather than imposing a particular gauge condition right from the beginning,
it is often advantageous to express the perturbations without specifying
any gauge. This then adds the flexibility of adopting different gauge
conditions at a much later stage, depending upon the nature of each
problem. Moreover, it becomes easy to relate results between various
gauge-dependent and gauge-invariant techniques. This approach has been termed
the \textit{gauge-ready} method. 

In this paper we apply the gauge-ready method to analyze 
perturbations in cosmological inflation driven by multicomponent
scalar fields. This paper is organized as follows. In Section \ref{formal}
we set up the equations describing multicomponent scalar fields with a
non-trivial field metric coupled non-minimally to Einstein gravity. We
then introduce a set of basis vectors, using Gram-Schmidt orthonormalization,
which enables us to disentangle multiple-field effects from single-fields ones.
The background equations, metric perturbations and the perturbed order
variables are presented in Section \ref{pertuniv}. Here we also discuss
briefly the issue of gauge transformations as applied to cosmological
perturbations. In Section \ref{scalar-pert} we introduce the gauge-ready
approach to cosmological perturbations in the multiple-field inflation
scenario. We present the perturbation equations in the gauge-ready form,
as well as in terms of gauge-invariant variables derived from the gauge-ready
equations. Slow-roll variables in the context of multicomponent inflation
are presented in Section \ref{solpert}. We proceed to apply canonical
quantization to the density perturbations. The solutions to the equations for
quantized perturbations are then derived to the first order in slow-roll.
A brief discussion of vector and tensor perturbations is also presented.
We conclude in Section \ref{discuss}.
\section{Preliminaries}
\labl{formal}
\subsection{The scalar fields}
\labl{scalarfield}
In this Section, we explain our notation and set up the basic equations
needed for our analysis.
As our starting point, we consider Einstein gravity  
coupled to an arbitrary
number of real scalar fields. We write the Lagrangean as
\equ{
\mathscr L &=&
       \sqrt{-g} \lh \frac {1}{2 \gksq} R -
       \half \der^\gm \Bgf \cdot \der_\gm\Bgf
       - V(\Bgf) \rh                         \nonumber \\
    &=&
      \sqrt{-g} \lh \frac {1}{2 \gksq} R -
      \half g^{\mu\nu} \der_\mu \Bgf^T \mx G \der_\nu \Bgf
      - V(\Bgf)  \rh . 
\labl{action}
}
Here $R$ is the scalar curvature, $\gk_0^2 \equiv 8 \gp G$, and we set
$c = 1$. For the scalar fields we use a vector notation,
$\Bgf \equiv (\gf^a)$,
where the indices $a,b,c,\ldots = 1,2,3,$ $\ldots, N$ 
label the $N$--components in field space.
Further, $g \equiv {\text{det}}(g_{\gm \gn}) $,
and $\gm, \gn, \ldots$ denote the spacetime indices. For repeated indices, 
the summation convention applies. The second quantity within the parentheses
of Eq.~(\ref{action}) represents a nonlinear sigma-model like non-minimal 
kinetic term. Such a kinetic term appears in various models of 
high-energy physics \cite{Lythrep}. Also $V(\Bgf)$ is an arbitrary scalar
potential.

Following the authors of \cite{Tent}, we can interpret the scalars $\Bgf$ as
coordinates $(\gf^a)$ on a real manifold $\cM$ induced with a symmetric
Riemannian metric $\mx G$ having components $G_{a b}$ in the field space. The 
field metric is chosen to be positive-definite so that the Hamiltonian is
bounded from below. The special case of minimally-coupled fields corresponds
to the situation $G_{a b} \equiv \gd_{a b}$.
From the components $G_{a b}$ we can define the
connection coefficients $\gG^a_{b c}$ in the usual manner,
\beq
\gG^a_{bc} = \half G^{ad}
\lh G_{bd,c} + G_{cd,b} - G_{bc,d} \rh.
\labl{connection}
\eeq
The curvature
tensor on $\cM$ is introduced in terms of the tangent vectors
$\vc{B}, \vc{C}, \vc{D}$:
\begin{widetext}
\beq
[\mx{R}(\vc{B},\vc{C})\vc{D}]^a \equiv
R^a_{\; bcd} \, B^b C^c \, D^d \equiv
\lh \gG^a_{bd,c} - \gG^a_{bc,d} + \gG^e_{bd} \gG^a_{ce}
- \gG^e_{bc} \gG^a_{de} \rh  B^bC^c \, D^d .
\labl{curve}
\eeq
\end{widetext}
For any two vectors $\vc{A}$ and $\vc{B}$, we define the inner product and
the norm as
\equ{
& & \vc{A} \cdot \vc{B} =
\vc A^\dag \vc B \equiv
\vc{A}^T \mx{G} \vc{B} = A^a G_{ab} B^b , \nonumber
\\
& & |\vc{A}| \equiv \sqrt{({\vc{A}\cdot\vc{A}})} ,
\labl{prod}
}
respectively. Here $\vc A^\dag$ is the cotangent vector such that
$(\vc A^\dag)_a \equiv A^b G_{ba}$. We also introduce the covariant
derivative $\nabla_a$ on $\cM$ acting upon a vector $\vc{A}$ as
\beq
\nabla_b A^a \equiv A^a_{\; ,b} + \gG^a_{bc} A^c ,
\labl{deriv1}
\eeq
while, the covariant derivative on $\vc{A}$ with respect to
the spacetime $x^\gm$ is
\beq
\cD_\gm A^a \equiv \der_\gm + \gG^a_{bc} \der_\gm \gf^b A^c .
\labl{deriv2}
\eeq
It should be noted that the covariant derivative reduces to the
ordinary derivative when it acts upon a scalar.

By varying the action (\ref{action}) with respect to $g_{\gm \gn}$
and $\Bgf$, we obtain the gravitational field equation,
\beq
\frac {1}{\gksq} G^\gm_{\; \nu} = T^\mu_{\;\nu}
= \der^\mu \Bgf \cdot \der_\nu \Bgf
- \gd^\mu_\nu \lh \half \der^\gl \Bgf \cdot \der_\gl \Bgf + V \rh ,
\labl{tmunu}
\eeq
and the equation of motion for the scalar fields,
\beq
g^{\mu\nu} \lh \cD_\mu \gd^\gl_\nu - \gG^\gl_{\mu\nu} \rh \der_\gl \Bgf
- \mx G\inv \Bnabla^T V = 0 ,
\labl{eqmot}
\eeq
where $G^\gm_{\; \gn}$ and $T^\gm_{\; \gn}$ are Einstein and
energy-momentum tensors.
 
It is often convenient in our analysis to represent the scalar fields 
as effective fluid quantities. We conclude this Section by covariantly
decomposing the energy-momentum tensor into fluid quantities using a
time-like four-vector $u^\gm$ normalized as $u^\gm u_\gm = -1$: 
\equ{
& & T_{\ga \gb} = \gm u_\ga u_\gb + p h_{\ga \gb} + q_\ga u_\gb +
q_\gb u_\ga + \gp_{\ga \gb},                \nonumber
\\
& & \gm \equiv T_{\ga \gb} u^\ga u^\gb ,
\quad
p \equiv \frac{1}{3}T_{\ga \gb}h^{\ga \gb} , \nonumber
\quad
q_\ga \equiv -T_{\gb \ggam} u^\gb h^\ggam_\ga ,  \nonumber
\\
& & \gp_{\ga \gb} \equiv T_{\ggam \gd}h^\ggam_\ga h^\gd_\gb - ph_{\ga \gb}.
\labl{fluid-decomp}
}
Here $\gm$, $p$, $q_\ga$, and $\gp_{\ga \gb}$ are the energy density,
pressure, energy flux, and anisotropic pressure, respectively;
$h_{\ga \gb} \equiv g_{\ga \gb}+u_\ga u_\gb$ is a projection
tensor based on $u_\ga$ vector, $q_\ga u^\ga = 0 = \gp_{\ga \gb}$,
and $\gp^\ga_\ga = 0$.
The decomposition given above is in the most general form. Indeed, for
a multicomponent scalar field, we have
\equ{
& & \gm = \half|\dot{\Bgf}|^2+V ,
\quad
p = \half|\dot{\Bgf}|^2-V , \nonumber
\\
& & q_\ga = 0 = \gp_{\ga \gb}.
\labl{scalar-fluid}
}
Equations (\ref{tmunu}) and (\ref{eqmot}) with (\ref{scalar-fluid}) provide 
the fundamental expressions required for describing cosmological inflation. 

\subsection{Basis vectors}
\labl{basis}
We continue our discussion of fields on a manifold by introducing
a set of basis vectors \cite{Tent}, which will prove to be useful
in our analysis. First note that an arbitrary tangent vector
$\vc{A}$ on $\cM$ can always be expanded in terms of a set of 
basis vectors $\{\vc{e}_{(a)}\}$ as $\vc{A} \equiv A^{(a)} \vc{e}_{(a)}$ 
with $\vc{e}_{(a)} \cdot \vc{e}_{(b)} = G_{ab}$.
However, a different set of
basis vectors generated using Gram-Schmidt orthonormalization
turns out to be more convenient.

From the vector $\Bgf$ we can construct a set of $N$ linearly 
independent vectors $\{\Bgf^{(1)},\Bgf^{(2)},\ldots,\Bgf^{(N)}\}$,
where,
\equ{
\Bgf^{(1)} \equiv \dot{\Bgf},
\quad
\Bgf^{(n)} \equiv \cDt^{(n-1)}\dot{\Bgf}\;\; 
\quad
(n \geq 2).
\labl{shorthand}
}
Let $\vc{e}_1 = \Bgf^{(1)}/{|\Bgf^{(1)}|}$ be the first unit vector
along the direction of the field velocity $\dot{\Bgf}$. Define the
second unit vector $\vc{e}_2$ to be along that part of the direction
of the field accelaration $\cDt \dot{\Bgf}$ which is normal 
to $\vc{e}_1$:
\beq
\vc{e}_2 = \frac{\Bgf^{(2)}-(\vc{e}_1 \cdot \Bgf^{(2)})\vc{e}_1} 
{|\Bgf^{(2)}-(\vc{e}_1 \cdot \Bgf^{(2)})\vc{e}_1|}.
\labl{e2}
\eeq
It is obvious from Eq.~(\ref{e2}) that 
$\vc{e}_1 \cdot \vc{e}_2 = 0$ by construction. 
A repetitive application of the Gram-Schmidt procedure then 
generates a set of mutually orthonormal vectors $\{\vc{e}_n\}$,
which span the same subspace as the vectors $\{\Bgf^{(n)}\}$.

Introducing the projection operators $\mx P_n$ and $\mx P^{\perp}_n$,
which project on $\vc{e}_n$ and on the subspace perpendicular
to $\{\vc{e}_1,\ldots,\vc{e}_n\}$ respectively, 
we may then write a general unit vector $\vc{e}_n$ as,
\beq
\vc{e}_n = \frac{\mx P^{\perp}_{n-1}\Bgf^{(n)}} 
{|\mx P^{\perp}_{n-1}\Bgf^{(n)}|}, 
\labl{en}
\eeq
where,
\equ{
\mx P_n = \vc{e}_n \vc{e}^{\dag}_n,
\quad
\mx P^{\perp}_n = \Id - \sum^{n}_{q=1}\mx P_q,
\quad
\mx P^{\perp}_0 \equiv \Id,
\labl{proj}
}
and we define,
\equ{
\mx P^{\parallel} \equiv \mx P_1 = 
\vc{e}_1 \vc{e}^{\dag}_1, 
\quad
\mx P^{\perp} \equiv \mx P^{\perp}_1 = \Id - \mx P^{\parallel}.
\labl{parperp1}
}
Note that when the denominator
in Eq.~(\ref{en}) vanishes, the corresponding basis vector does not
exist.

Using the fact that $\mx P^{\parallel}+\mx P^{\perp} \equiv \Id$, we can
decompose any vector $\vc{A}$ in directions parallel and perpendicular
to the field velocity:
\equ{
\vc{A} &=& \vc{A}^{\parallel} + \vc{A}^{\perp}
\equiv (\mx P^{\parallel}+ \mx P^{\perp})\vc{A} \nonumber
\\
&=& \vc{e}_1(\vc{e}_1 \cdot \vc{A})+
\vc{e}_2(\vc{e}_2 \cdot \vc{A}).
\labl{parperp2}
}
For the special case of just one field, $\vc{e}_1$ by definition simply
reduces to the normalized scalar $\gf^{(1)}/{|\gf^{(1)}|}$. Hence, from
Eq.~(\ref{e2}), $\vc{e}_2$ vanishes identically, and so do all other basis
vectors. Thus the decomposition (\ref{parperp2}) enables us to distinguish
between single-field contributions, where only $\vc{e}_1$ survives,
from multiple-field ones. 
\section{The perturbed universe}
\labl{pertuniv}
\subsection{Metric perturbations}
\labl{pert}
We know that the observed Universe is not perfectly homogeneous and
isotropic. Assuming that the inhomogeneities are small
enough, we can then treat the deviations using perturbation theory.
In this paper we consider linear perturbations of the homogeneous 
and isotropic cosmological space-time described by the 
Friedmann-Robertson-Walker ( FRW ) model.
We choose the line-element to be 
\equ{
d s^2 &=& -a^2 \lh 1 + 2 A \rh d \get^2 -2 a^2 B_i d x^i \nonumber \\
      & & + a^2 \lh g^{(3)}_{i j} + 2 C_{i j} \rh d x^i d x^j ,
\labl{metric}
}
where $a(t)$ is the scale factor, $dt \equiv a d\get$,
and indices $i,j,\ldots$, run from $1$ to $3$ labelling the spatial
components. The perturbed order variables
$\;A(t, \vc{x})$, $B_i(t, \vc{x})$, and
$C_{i j}(t, \vc{x})$ are based on the metric $g^{(3)}_{i j}$
of the 3-surfaces of constant curvature $K = 0, \pm1$.
These are general functions of space-time which characterize
the linear cosmological perturbations. Here $t$ and
$\get$ are the comoving and conformal times respectively. We denote a
derivative with respect to comoving time by $\;\dot{} \equiv \der_t$ and
one with respect to conformal time by$\;\prime \equiv \der_\get$. The
Hubble parameters in terms of comoving and conformal times are defined
as $H = \dot a /a$ and $\cH = a' /a = aH$.

As is evident from Eq.~(\ref{metric}), the metric is decomposed into
a background part, plus a perturbation. Correspondingly we can 
decompose the scalar field as
\beq
\Bgf(t, \vc{x})  = \bar{\Bgf}(t) + \Bgd \Bgf(t, \vc{x}) ,
\labl{phidecomp}
\eeq 
where the perturbation $\Bgd\Bgf \equiv (\gd \gf^a)$ 
is a tangent vector on $\cM$,
while the energy-momentum tensor is decomposed as
\equ{
& & T^0_{\; 0} = - \gm \equiv -(\bar{\gm}+\gd \gm) , \nonumber
\\
& & T^0_{\; i} = \frac{1}{a}[q_i + (\gm + p)u_i]
\equiv (\gm + p)v_i , \nonumber
\\
& & T^i_{\; j} = p \gd^i_j + \gp^i_j
\equiv (\bar p + \gd p)\gd^i_j + \gp^{(3)i}_{\;\;\;\;\; j}.
\labl{tmndecomp}
}
The barred entities denote background variables. For notational
simplicity we shall ignore the overbars unless required.
In Eq.~(\ref{tmndecomp}),
$v_i$ is the frame-independent flux variable, and
$v_i$, $\gp^{(3)i}_{\;\;\;\;\; j}$ are based on
$g^{(3)}_{i j}$. 

From Eqs.~(\ref{tmunu}) and
(\ref{eqmot}), the equations for the background can be written as
\equ{
& &H^2 = \frac{1}{3} \gksq  \gm - \frac{K}{a^2}
= \frac{1}{3} \gksq \lh \half |\dot  {\Bgf}|^2 + V \rh - \frac{K}{a^2} ,
\labl{b1}
\\
& &\dot H = - \half \gksq \lh \gm + p \rh + \frac{K}{a^2}
= - \half \gksq \; |\dot {\Bgf}|^2 + \frac{K}{a^2} ,
\labl{b2}
\\
& &R = 6 \lh 2 H^2 + \dot H + \frac{K}{a^2} \rh ,
\labl{b3}
\\
& &\cDt \dot {\Bgf} + 3 H \dot {\Bgf} +
\mx G\inv \Bnabla^T V = 0 ,
\labl{b4}
\\
& &\dot \gm + 3 H \lh \gm + p \rh = 0.
\labl{b5}
}
Taking the $G^0_0$ and
$G^i_i- 3G^0_0$ components of Eq.~(\ref{tmunu}) yield
Eqs.~(\ref{b1}) and (\ref{b2}) respectively,
while Eq.~(\ref{eqmot}) leads to Eq.~(\ref{b4}).
Equation (\ref{b5}) follows from the conservation of 
the energy-momentum tensor. We shall ignore the cosmological
constant $\gL$ in our work; nevertheless it can be easily
included by making the replacements $\gm \ra \gm + \gL/\gksq$
and $p \ra p - \gL/\gksq$. Note that we have explicitly 
retained $K ( = 0, \pm 1)$, and only at a later stage shall we 
set $K = 0$.

\subsection{Scalar, vector and tensor decompositions}
\labl{svt}
In order to make further progress, it is customary to decompose
the perturbed order variables into scalar-, vector-, and tensor-type
perturbations. To the linear order, each of these three perturbations
decouple from one another and evolve independently. Accordingly, the
metric perurbation variables $A(t, \vc{x})$, $B_i(t, \vc{x})$, and
$C_{i j}(t, \vc{x})$ may be decomposed as
\equ{
& & A \equiv \ga,     \nonumber
\quad
B_i \equiv \gb_i + B^{(v)}_i, \nonumber
\\
& & C_{i j} \equiv g^{(3)}_{i j}\gvf + \ggam_{, i|j} + C^{(v)}_{(i|j)}
+ C^{(t)}_{i j}.
\labl{svtdec}
} 
In this and the following, the superscripts $(s)$, 
$(v)$ and $(t)$ will indicate the scalar-, vector- 
and tensor-type perturbed order variables. The vertical bar represents
a covariant derivative with respect to $g^{(3)}_{i j}$ and the round 
brackets in the subscript imply symmetrization of the indices. The scalar 
metric perturbations are then given by $\ga$, $\gb$, $\ggam$ and $\gvf$. The
transverse-type vector perturbations $B^{(v)}_i$ and $C^{(v)}_i$ satisfy
$B^{(v) i}_{\;\;\;\;\;| i} = 0 = C^{(v) i}_{\;\;\;\;\;| i}$ 
while the tensor-type perturbation
$C^{(t)}_{i j}$ is 
transverse-traceless $( C^{(t) i}_{\;\;\;\;\;i} = 0 
= C^{(t) j}_{\;\;\;\;\;i|j} )$. Both the vector and tensor perturbed order
variables are based on $g^{(3)}_{i j}$. 
We define $\gD$ as a comoving three-space Laplacian, and introduce
the following combinations of the metric variables,
\equ{
& & \gc \equiv a ( \gb + a \dot{\ggam} ) , \nonumber
\quad
\gk \equiv 3 ( H \ga - \dot{\gvf} - \frac{\gD}{a^2} \gc ) , \nonumber
\\
& & \gPs^{(v)} \equiv B^{(v)} + a \dot C^{(v)}.
\labl{combi}
}

It is convenient to separate the temporal and spatial aspects of the
perturbed order variables by expanding them in terms of 
harmonic eigenfunctions
$\cQ^{(s,v,t)}(\vc{k};\vc{x})$ of the generalized Helmholtz equation
\cite{Bardeen80, Kodamasasaki}: 
\equ{
& & \cQ^{(s)\; |i}_{, i} \equiv - k^2 \cQ^{(s)}, \nonumber
\quad
\cQ^{(s)}_i \equiv - \frac{1}{k}\cQ^{(s)}_{, i},
\\
& & \cQ^{(s)} \equiv \frac{1}{k^2}\cQ^{(s)}_{,i|j} + \nonumber
\frac{1}{3}g^{(3)}_{i j}\cQ^{(s)},
\\
& & \cQ^{(v)\; |i}_{i, j} \equiv - k^2 \cQ^{(v)}_i, \nonumber
\quad
\cQ^{(v)}_{i j} \equiv - \frac{1}{k}\cQ^{(v)}_{(i|j)},
\quad
\cQ^{(v)\; |i}_i \equiv 0,
\\
& & \cQ^{(t)\; |k}_{ij, k} \equiv - k^2 \cQ^{(t)}_{ij}, \nonumber
\quad
\cQ^{(t)}_{ij} \equiv \cQ^{(t)}_{ji},
\\
& & \cQ^{(t)\; |j}_{ij} \equiv 0 \equiv \cQ^{(t) i}_{\;\;\;\;i}. 
\labl{harmonic}
}
Here $\vc{k}$ is the wave vector in Fourier space and $k = |\vc{k}|$.
We can then write the scalar-type perturbed order variables as
$\ga(t, \vc{x}) \equiv \ga(t, \vc{k})\cQ^{(s)}(\vc{k};\vc{x})$, with 
similar expressions for $\gb$, $\ggam$ and $\gvf$. The vector- and tensor-type
perturbations are expanded as $B^{(v)}_i \equiv B^{(v)}\cQ^{(v)}_i$,
$C^{(v)}_i \equiv C^{(v)}\cQ^{(v)}_i$, and 
$C^{(t)}_{i j} \equiv C^{(t)}\cQ^{(t)}_{i j}$. In each of these harmonic 
expansions, a summation over the modes of the eigenfunctions is implied.
In particular, the perturbed scalar fields have the expansion
\beq
\Bgd \Bgf(t, \vc{x}) \equiv \Bgd \Bgf(t, \vc{k})\cQ^{(s)}(\vc{k};\vc{x}).
\labl{phiharmonic}
\eeq 
From Eq.~(\ref{phiharmonic}) we derive the important conclusion that
the scalar fields on their own can generate only scalar-type perturbations.
We also note that to the
linear order in perturbations, the form of the equations in configuration
space are identical to the corresponding ones in Fourier space. 
Consequently, for maintaining notational ease, we do not distinguish
between the two.

In a similar spirit, the fluid variables $v_i$ 
and $\gp^{(3)i}_{\;\;\;\;\; j}$ can be expanded in terms of the harmonics 
as
\equ{
& & v_i \equiv v^{(s)}\cQ^{(s)}_i + v^{(v)}\cQ^{(v)}_i, \nonumber
\\
& & \gp^{(3)i}_{\;\;\;\;\; j} \equiv \gp^{(s)}\cQ^{(s) i}_{\;\;\;\; j}
    + \gp^{(v)}\cQ^{(v) i}_{\;\;\;\; j}
    + \gp^{(t)}\cQ^{(t) i}_{\;\;\;\; j};
\labl{fluidharmonic}
}
while the energy-momentum tensor in Eq.~(\ref{tmndecomp}) has the expansion
\equ{
& & T^0_{\; 0} = - \gm \equiv -(\bar{\gm}+\gd \gm) , \nonumber
\\
& & T^0_{\; i} = -\frac{1}{k}(\gm + p)v^{(s)}_{, i}
    + (\gm + p)v^{(v)}\cQ^{(v)}_i ,   \nonumber
\\
& & T^i_{\; j} = (\bar p + \gd p)\gd^i_j
    + \gp^{(s)}\cQ^{(s) i}_{\;\;\;\; j}
    + \gp^{(v)}\cQ^{(v) i}_{\;\;\;\; j}         
    + \gp^{(t)}\cQ^{(t) i}_{\;\;\;\; j}.
\nonumber \\    
\labl{tmnharmonic}
}

For a Universe having the matter sector composed
exclusively of scalar fields, the quantity $\gp^{(3)i}_{\;\;\;\;\; j}$
in Eq.~(\ref{tmndecomp}) vanishes identically. We then have
to the perturbed order,  
\equ{
& & \gd \gm = \dot{\Bgf}\cdot\cDt\gd\Bgf - \ga |\dot{\Bgf}|^2 
+ \Bnabla V \cdot \gd \Bgf,
\labl{delmu}
\\ 
& & \gd p = \dot{\Bgf}\cdot\cDt\gd\Bgf - \ga |\dot{\Bgf}|^2 
- \Bnabla V \cdot \gd \Bgf,
\labl{delp}
\\
& & (\gm + p)v \frac{a}{k}= \dot{\Bgf} \cdot \gd \Bgf,
\labl{mup} 
}
where we have written $v \equiv v^{(s)}$ for simplcity.
It is also convenient to decompose $\gd p$ into an adiabatic part
$c^2_s \gd \gm$, and an entropic perturbation $e$:
\beq
\gd p = c^2_s \gd \gm + e,
\labl{ad-ent}
\eeq
where $c^2_s \equiv \dot{p}/\dot{\gm}$ may be interpreted as an effective 
sound velocity.  
\subsection{Gauge transformations}
\labl{gaugetrans}
As mentioned previously, our Universe shows departures from ideal
homogeneity and isotropy. When the deviations are small enough,
one can choose a fictitious background geometry and consider
perturbations about it using infinitesimal coordinate transformations.
The change in correspondence between the background and perturbed 
space-times represented by coordinate shifts is called
a \textit{gauge transformation}.
Now, since relativistic gravity is invariant under coordinate
transformations, the perturbations are not unique. There are different
ways of mapping between the background and perturbed parts. This leads to
what is called gauge degrees of freedom in the context of cosmological
perturbations. 

Below we briefly summarize the transformation properties of various
quantities to the linear order in perturbations \cite{Mukhanovetal,
Bardeen80, Kodamasasaki}. Under a coordinate shift
$\tx^\gm = x^\gm + \gx^\gm$, the scalars, vectors and tensors transform as
\equ{
& & \tilde{\cA}(\tx^\gl) = \cA(x^\gl), \nonumber 
\quad
\tilde{\cA}_\gm(\tx^\gl)= \frac{\der x^\gs}{\der \tx^\gm}
\cA_\gs(x^\gl), \nonumber
\\
& & \tilde{\cA}_{\gm \gn} (\tx^\gl)
= \frac{\der x^\gs}{\der \tx^\gm} \frac{\der x^\gt}{\der \tx^\gn} 
\cA _{\gs \gt} (x^\gl),
\labl{gaugeshift}
}
so that
\equ{
& &\tilde{\cA}(x^\gl) = \cA(x^\gl) - \cA_{, \gs}\gx^\gs, \nonumber
\\
& &\tilde{\cA}_\gm(x^\gl) = \cA_\gm(x^\gl)- \cA_{\gm, \gs}\gx^\gs
- \cA_\gs \gx^\gs_{, \gm}, \nonumber
\\
& &\tilde{\cA}_{\gm \gn} (x^\gl)
= \cA_{\gm \gn} (x^\gl) - \cA_{\gm \gn, \gs}\gx^\gs
- 2 \cA_{\gs ( \gn}\gx^\gs_{, \gm )}.
\labl{infgaugeshift}
}
Writing the temporal part of $\gx_\gm$ as $\gx^0 = a\inv \gx^t$ and
decomposing the spatial part as $\gx_i = a \inv \gx_{, i} + \gx^{(v)}_i$,
where $\gx^{(v)}_i$ is based on $g^{(3)}_{i j}$
satisfying $\gx^{(v) i}_{\;\;\;\; |i} = 0$,
we find from Eq.~(\ref{infgaugeshift}) that
the metric and matter variables transform to linear order as:
\equ{
& & \tga = \ga - \dot{\gx^t}, \nonumber
\quad \tgb = \gb - \frac{1}{a}\gx^t + a \lh \frac{\gx}{a} \rh^{.},
\\
& & \tgg = \ggam - \frac{1}{a} \gx, \nonumber
\quad
\tgvf = \gvf - H\gx^t,
\quad
\tgc = \gc - \gx^t,
\\
& & \tgk = \gk + \lh 3 \dot H + \frac{\gD}{a^2} \rh, \nonumber
\quad
\tv = v - \frac {1}{a} \gx^t,
\\
& & \gd \tgm = \gd \gm - \dot \gm \gx^t, \nonumber
\quad
\gd \tp = \gd p - \dot p \gx^t,
\quad
\Bgd \tilde{\Bgf} = \Bgd \Bgf - \dot{\Bgf} \gx^t,
\\
& & \tB^{(v)}_i = B^{(v)}_i + a \dot{\gx}^{(v)}_i, \nonumber
\quad
\tC^{(v)}_i = C^{(v)}_i - \gx^{(v)}_i,
\\
& & \tv^{(v)} = v^{(v)}, \nonumber
\quad
\tgPs^{(v)} = \gPs^{(v)},
\\
& & \tgp^{(s, v, t)} = \gp^{(s, v, t)}, 
\quad
\tC^{(t)}_{i j} = C^{(t)}_{i j}.
\labl{transforms}
}

It is immediately obvious from Eq.~(\ref{transforms}) that
the tensor-type perturbations are gauge-invariant. For the
special case of scalar-type perturbations to the linear order, 
fixing the temporal
part $\gx^t$ of the gauge transformation leads to different
gauge conditions. Table~\ref{gauges} summarizes some common temporal
gauges.

\begin{table}
\begin{ruledtabular}
\begin{tabular}{lll}
\textbf{Gauge} & \textbf{Gauge} & \textbf{Temporal} \\
& \textbf{condition} & \textbf{component} \\
\hline
Synchronous gauge & $\ga \equiv 0$ & $\gx^t(x)$ \\
Zero-shear gauge  & $\gc \equiv 0$ & $\gx^t = 0$ \\
Comoving gauge & $v/k \equiv 0$ & $\gx^t = 0$ \\
Uniform-curvature gauge & $\gvf \equiv 0$ & $\gx^t = 0$ \\
Uniform-expansion gauge & $\gk \equiv 0$ & $\gx^t = 0$ \\
Uniform-field gauge & $\Bgd \Bgf \equiv 0$ & $\gx^t = 0$ \\
Uniform-density gauge & $\gd \gm \equiv 0$ & $\gx^t = 0$ \\
Uniform-pressure gauge & $\gd p \equiv 0$ & $\gx^t = 0$ \\
\end{tabular}
\end{ruledtabular}
\caption{Common gauge conditions}
\labl{gauges}
\end{table}
From the first equation in (\ref{transforms}), we see that the
gauge transformation of $\ga$ involves the term $\dot{\gx^t}$. Therefore
the synchronous gauge condition $\ga \equiv 0$ fixes $\gx^t$ upto a 
constant of integration, leaving a spatially varying 
residual gauge mode $\gx^t(\vc{x})$.
For the remaining gauges in Table~\ref{gauges}, the temporal gauge mode
is completely determined.

We conclude by giving a few examples of gauge-invariant
combination of variables:
\equ{
& & \gvfc \equiv \gvf - H \gc, \nonumber
\quad
\gac \equiv \ga - \dot \gc,
\quad
v_\gc \equiv v - \frac{k}{a}\gc,
\\
& & \gd \gm_\gc \equiv \gd \gm - \dot{\gm} \gc,  \nonumber
\quad
\gd p_\gc \equiv \gd p - \dot{p} \gc,
\\
& & \Bgd \Bgfc \equiv \Bgd \Bgf - \dot{\Bgf}\gc, \nonumber
\quad
\Bgd \Bgf_\gvf \equiv \Bgd \Bgf - \frac{\dot{\Bgf}}{H}\gvf 
\equiv - \frac{\dot{\Bgf}}{H}\gvf_{\Bgd \Bgf},
\\
& &  \gvf_v \equiv \gvf - \frac{aH}{k}v,
\quad
\gd \gm_v \equiv \gd \gm - \frac{a}{k}\dot \gm v.
\labl{givars}
}
Thus, in the zero-shear gauge, also known as the longitudinal, or
conformal Newtonian gauge, $\gc \equiv 0$ is the gauge
condition. We then have from Eq.~(\ref{givars}), $\gvfc \equiv \gvf$,
$\gac \equiv \ga$, and $\Bgd \Bgfc \equiv \Bgd \Bgf$. Similarly,
in the uniform-curvature gauge, it follows that
$\Bgd \Bgf_\gvf \equiv \Bgd \Bgf$ which in turn is equivalent to
$-(\dot{\Bgf}/H)\gvf_{\Bgd \Bgf}$ in the 
uniform-field gauge. In the notation of \cite{Mukhanovetal}, our
$\gac$ and $\gvfc$ correspond to their $\gF$ and $-\gPs$ respectively.  
\section{Scalar perturbations in multiple-field inflation}
\labl{scalar-pert}
\subsection{Perturbation equations in the gauge-ready form}
\labl{gauge-ready}
We shall now briefly discuss the gauge-ready approach introduced
in \cite{Hwang1,Hwang2,Hwang3,Hwang4}. As is well known in the theory
of cosmological perturbations, a judicious choice of gauge conditions 
often simplifies the mathematical structure of a particular problem. 
For example, density
perturbations with hydrodynamical fluids are most conveniently treated
using the comoving gauge, while the uniform-curvature gauge simplifies
the analysis of perturbations due to minimally coupled scalar fields.
Since, in general, we do not know the optimal gauge condition beforehand,
it becomes advantageous to express the perturbations without imposing
a specific temporal gauge condition. In other words, we write the 
governing equations in the \textit{gauge-ready} form, which would give us
the freedom to choose different gauge conditions, as adapted to the 
problem, at a later stage in the calculations. Once the temporal gauge mode 
is completely fixed so that no further gauge degrees of freedom are left,
the resulting variables would then be gauge-invariant.
Moreover, when a solution in a particular gauge  
is known, we can then easily derive the corresponding solution in 
other gauges, as well as in gauge-invariant forms. This is the basic
concept of the gauge-ready method.

To implement this gauge-ready strategy, it is most convenient to derive 
the perturbed set of equations from the (3+1) ADM \cite{ADM}, and the
(1+3) covariant \cite{covariant} formulations of Einstein gravity. A
complete set of these equations may be found in the Appendix
of Ref.~\cite{Hwang1}. In this Section we write the equations 
for scalar-type perturbations in the gauge-ready form.

\noindent Definition of $\gk$:
\beq
\dot{\gvf}= H \ga - \frac{1}{3} \gk + \frac{1}{3} \frac{k^2}{a^2} \gc.
\labl{g1}
\eeq

\noindent ADM energy constraint ($G^0_0$ component of the field equation):
\beq
- \frac{k^2-3K}{a^2} \gvf +H \gk = -\half \gksq \gd \gm.
\labl{g2}
\eeq 

\noindent ADM momentum constraint ($G^0_i$ component):
\beq
\gk - \frac{k^2-3K}{a^2}\gc = \frac{3}{2}\gksq(\gm+p) \frac{a}{k} v.
\labl{g3}
\eeq

\noindent ADM propagation($G^i_j-\frac{1}{3}\gd^i_jG^k_k$ component):
\beq
\dot \gc+H\gc-\ga-\gvf=\gksq \frac{a^2}{k^2}\gp^{(s)}.
\labl{g4}
\eeq

\noindent Raychaudhuri equation ($G^i_i-G^0_0$ component):
\beq
\dot{\gk}+2H\gk+\lh 3\dot{H}-\frac{k^2}{a^2}\rh \ga=\half \gksq (\gd \gm
+3\gd p).
\labl{g5}
\eeq

\noindent Equation of motion for scalar fields:
\equ{
\lh \cDt^2+3H\cDt-\frac{\gD}{a^2}+\mx M^2 \rh \Bgd \Bgf 
&=& \lh \dot{\ga}-3\dot{\gvf}-\frac{\gD}{a^2}\gc \rh \dot{\Bgf} \nonumber
\\
&-& 2\ga \mx G \inv \Bnabla ^T V.
\labl{g6}
}

\noindent Energy conservation:
\beq
\gd\dot{\gm}+3H(\gd \gm+\gd p)=(\gm + p)\lh \gk-3H\ga - \frac{k}{a}v \rh. 
\labl{g7}
\eeq

\noindent Momentum conservation:
\equ{
& & \frac{[a^4(\gm + p)v]^{\;\dot{}}}{a^4(\gm + p)} \nonumber
\\
& & =\frac{k}{a}\left[\ga + \frac{1}{\gm + p}\lh \gd p - \frac{2}{3} 
\frac{k^2-3K}{k^2}\gp^{(s)} \rh \right].
\labl{g8}
}

In the above equations, $\gd \gm$ and $\gd p$ are given by Eqs.~(\ref{delmu})
and (\ref{delp}) respectively, while 
\beq
\mx M^2 = \mx G \inv \Bnabla^T \Bnabla V - \mx R (\dot{\Bgf},\dot{\Bgf}).
\labl{M-def}
\eeq
Note that these equations are valid for any $K$, and for a scalar field,
$\gp^{(s)}=0$.

Equations (\ref{g1})-(\ref{g8}), together with the background equations
(\ref{b1})-(\ref{b5}), and the perturbed order variables 
for the scalar fields (\ref{delmu})-(\ref{mup}),
provide a complete set of equations for analyzing scalar-type cosmological
perturbations with multicomponent scalar
fields. As we have not chosen a specific gauge so far,
Eqs.~(\ref{g1})-(\ref{g8}) are therefore in the gauge-ready form. This
allows us to impose any one of the available temporal gauge conditions,
which would then fix the temporal gauge mode completely, leading to 
gauge-invariant variables.
\subsection{Perturbation equations using gauge-invariant variables}
\labl{gipert}
In order to illustrate the gauge-ready method, we derive some useful
expressions in terms of the gauge-invariant variables introduced in
Section \ref{gaugetrans}.

From Eqs.~(\ref{g2}) and (\ref{g3}) we obtain
\beq
\frac{k^2-3K}{a^2}\gvfc = \half \gksq \gd \gm_v. 
\labl{gi1}
\eeq 
Eq.~(\ref{g4}) can be written as
\beq 
\gac + \gvfc = -\gksq \frac{a^2}{k^2}\gp^{(s)}.
\labl{gi2}
\eeq
Eqs.~(\ref{g3}),(\ref{g4}) and (\ref{g1}) lead to
\beq
\dot{\gvf}_\gc-H\gac=-\half \gksq(\gm+p)\frac{a}{k}v_\gc,
\labl{gi3}
\eeq
Eqs.~(\ref{g7}),(\ref{g8}) with (\ref{g3}) yield
\equ{
& & \gd \dot{\gm}_v + 3H\gd \gm_v \nonumber
\\
& & = -\frac{k^2-3K}{a^2}\left[(\gm + p)\frac{a}{k}v_\gc + 
2H \frac{a^2}{k^2}\gp^{(s)}\right],
\labl{gi4}
}
while Eqs.~(\ref{g4}) and (\ref{g8}) give
\equ{
& & \dot{v}_\gc + Hv_\gc       \nonumber
\\
& & =\frac{k}{a}\left[\gac + \frac{\gd p_v}{\gm + p}-
\frac{2}{3}\frac{k^2-3K}{a^2}\frac{\gp^{(s)}}{\gm + p}\right].
\labl{gi5}
}
Combining Eqs.~(\ref{gi1})-(\ref{gi5}) we can derive
\equ{
& & \ddot{\gvf}_\gc + (4+3c^2_s)H\gvfc - c^2_s\frac{\gD}{a^2} \gvfc \nonumber
\\
& & +\left[(\gm c^2_s-p) \nonumber
-2(1+3c^2_s)\frac{K}{a^2}\right]\gvfc
\\
& & =-\half \gksq \lh e-\frac{2}{3}\gp^{(s)} \rh \nonumber
\\
& & -\half \gksq \frac{\gm +p}{H}\lh \frac{2H^2}{\gm +p}
\frac{a^2}{k^2}\gp^{(s)}\rh,
\labl{gi6}
}
where we used Eq.~(\ref{ad-ent}). For the explicit forms of the 
gauge-invariant variables used in these equations, see Eq.~(\ref{givars}).
From Eq.~(\ref{gi2}) we can draw the important conclusion that, for 
scalar-fields, $\gac=-\gvfc$, since $\gp^{(s)}=0$. Using this result
the equation of motion for scalar fields becomes
\equ{
\lh \cDe^2+2H\cDe-\gD +a^2\mx M^2 \rh \Bgd \Bgfc 
= -4\gvf'_\gc \Bgf'    \nonumber
\\
+ 2a^2\gvfc \mx G \inv \Bnabla ^T V.
\labl{gifield}
}
It is also convenient to re-write Eqs.~(\ref{gi3}) and (\ref{gi6}) 
for the case of scalar fields as
\equ{
\gvf'_\gc+\cH \gvfc = -\half \gksq \Bgf' \cdot \Bgd \Bgfc,
\labl{constraint}
\\
\gvf''_\gc + 6\cH \gvf'_\gc - \gD \gvfc 
+ 2\left[\cH'+2(\cH^2-K)\right]\gvfc \nonumber
\\
=\gksq a^2 \Bnabla V \cdot \Bgd \Bgfc,
\labl{gipot}
}
where we used Eq.~(\ref{mup}), and the relations
\equ{
& & e=\gd p - c^2_s\gd \gm = \gd p_\gc - c^2_s \gd \gm_\gc, \nonumber
\\
\quad
& & (1 - c^2_s)\gd \gm_\gc - e = \gd \gm_\gc-\gd p_\gc 
= \Bnabla V \cdot \Bgd \Bgfc.
\labl{subsidiary}
}
Eq.~(\ref{constraint}) is often called the \textit{constraint} equation.
Note that we have written Eqs.~(\ref{gifield})-(\ref{gipot})in conformal
time. These equations contain most of the physics related to inflationary
cosmological perturbations. They are expressed in terms of gauge-invariant
forms of the variables, and from the discussion at the end of 
Section \ref{gaugetrans}, we see that they retain the same algebraic forms
in the zero-shear gauge.

We end this Section by expressing Eq.~(\ref{gipot}) in a different way.
Observe that according to Eq.~(\ref{parperp2}), $\Bgd \Bgfc$ may be 
decomposed into components parallel and perpendicular to the field 
velocity, $\Bgd \Bgfc = \Bgd \Bgf^{\parallel}_\gc + \Bgd \Bgf^{\perp}_\gc$.
Using the background equation (\ref{b4}), the constraint 
equation (\ref{constraint}), and the fact 
that $|\Bgf'|'|\Bgf'| = (\cDe \Bgf') \cdot \Bgf'$, we can write 
Eq.~(\ref{gipot}) as
\equ{
\gvf''_\gc &+& 2\lh \cH - \frac{|\Bgf'|'}{|\Bgf'|} \rh \gvf'_\gc \nonumber
\\
&+& 2\left[\lh \cH' - \cH \frac{|\Bgf'|'}{|\Bgf'|}\rh - 2K\right]\gvfc 
-\gD\gvfc                                                    \nonumber
\\
&=& -\gksq (\cDe \Bgf') \cdot \Bgd \Bgf^\perp.
\labl{newpot}
}
Following our discussion in Section \ref{basis}, we know that the
perpendicular component of field perturbation vanishes when there is 
only one field. In this case, the right hand side of Eq.~(\ref{newpot})
vanishes, and the resulting equation is well known in the theory of 
single field inflationary perturbations \cite{Mukhanovetal}.

\section{Solutions of the perturbation equations}
\labl{solpert}
\subsection{Slow-roll variables}
\labl{slowroll}
To proceed further with our analysis, we make the 
\textit{assumption} that the Universe has undergone 
inflation to complete flatness, so that henceforth we can set $K=0$.

We introduce the functions,
\equ{
\ge(\Bgf) \equiv -\frac{\dot{H}}{H^2},
\quad
\Bget(\Bgf) \equiv \frac{\Bgf^{(2)}}{H|\dot{\Bgf}|},
\labl{slrl}
}
known as the \textit{slow-roll} variables. Using Eq.~(\ref{parperp2}),
$\Bget$ is decomposed into parallel and perpendicular components:
\beq
\getp = \vc{e}_1 \cdot \Bget 
= \frac{\cDt\dot{\Bgf}\cdot\dot{\Bgf}}{H|\dot{\Bgf}|^2},
\quad
\getn = \vc{e}_2 \cdot \Bget
= \frac{|(\cDt \dot{\Bgf})^\perp|}{H|\dot{\Bgf}|}.
\labl{etacomp}
\eeq

The standard slow-roll assumptions are
\beq
\ge = O(\gz),
\quad
\getp = O(\gz),
\quad
\getn = O(\gz),
\labl{slrlassump}
\eeq
for some small parameter $\gz$, with $\ge$, $\sqrt\ge\getp$ and
$\sqrt\ge\getn$ much smaller than unity. 
If in an expansion in 
slow-roll variables we neglect terms of order $O(\gz^2)$, we claim
that expansion to be of first order in slow-roll. Thus terms with
$\ge^2$, $\ge\getp$, etc. are of second order.
Note that the
definitions (\ref{slrl}) remain valid irrespective of the
slow-roll assumptions.

We present some useful relations involving the slow-roll variables:
\equ{
& & \cH'=\cH^2(1-\ge), \nonumber
\quad
\frac{|\Bgf'|'}{|\Bgf'|}=\cH(1+\getp), 
\\
& & \cDe \Bgf'=\cH |\Bgf'|(\Bget + \vc{e}_1)=
\gk\inv_0\sqrt{2}\cH^2\sqrt{\ge}(\Bget+\vc{e}_1), \nonumber
\\
& & \cH^2 \ge=\half \gksq |\Bgf'|^2, 
\quad
\ge'=2\cH\ge(\ge+\getp).
\labl{useful}
}
\subsection{Analysis using gauge-invariant variables}
\labl{gisol}
In order to solve the system of perturbation equations (\ref{gifield}),
(\ref{constraint}) and (\ref{newpot}), we shall find it convenient to
introduce the variables
\equ{
& &\Bq=a\lh\Bgd\Bgfc-\frac{\Bgf'}{\cH}\gvfc \rh
=a\lh\Bgd\Bgf-\frac{\Bgf'}{\cH}\gvf \rh,
\labl{sasaki1}
\\
& &u=-\frac{a}{|\Bgf'|}\gvfc,
\labl{sasaki2}
} 
where the second equality for $\Bq$ in Eq.~(\ref{sasaki1}) follows from
Eq.~(\ref{givars}). Indeed, $\Bq$ is  gauge-invariant, and is a natural
generalization of the single field Sasaki-Mukhanov 
variable \cite{Mukhanovetal}.

Using the slow-roll variables (\ref{slrl}), together with some of the 
relations (\ref{useful}), the constraint equation (\ref{constraint})
can be written in terms of $\Bq$ as
\beq
\gvfc'+\cH(1+\ge)\gvfc=-\half\gksq\Bgf'\cdot\frac{\Bq}{a},
\labl{constq}
\eeq

From Eq.~(\ref{gifield}), the scalar field perturbations satisfy
\beq
\cDe^2 \Bq -(\gD - \cH^2\BgO)\Bq=0,
\labl{fieldeq}
\eeq
where
\beq
\BgO = \frac{a^2\mx M^2}{\cH^2}-(2-\ge)\openone
-2\ge\lh(3+\ge)\mx P^{\parallel}
+\vc{e}_1\Bget^\dag+\Bget\vc{e}^\dag_1\rh,
\labl{omega}
\eeq
and use has also been made of Eq.~(\ref{constq}).
The corresponding Lagrangean $\mathscr L$ follows from Eq.~(\ref{fieldeq}):
\equ{
S&=&\int \mathscr L \sqrt{g^{(3)}}d\get d^3\vc{x} \nonumber
\\
&=&\half \int
\lh\cDe\Bq^\dag\cDe\Bq+\Bq^\dag(\gD-\cH^2\BgO)\Bq\rh\sqrt{g^{(3)}}
d\get d^3\vc{x}.
\nonumber \\
\labl{fieldaction}
}
Here $g^{(3)}$ is the determinant of the metric $g^{(3)}_{i j}$ of the
3-surfaces of constant curvature $K=0$, see below Eq.~(\ref{metric}).

The equation of motion for $u$ is obtained by substituting
its definition (\ref{sasaki2}) into Eq.~(\ref{newpot}):
\equ{
& & u''-\gD u-\frac{\gth''}{\gth}u =\gksq \cH \getn q_2,
\quad
q_2\equiv \vc{e}_2 \cdot \Bq, \nonumber
\\
& & \gth \equiv \frac{\cH}{a|\Bgf'|}=\frac{\gk_0}{\sqrt2}\frac{1}{a\sqrt \ge}.
\labl{ueq}
}

For later use, we also express Eq.~(\ref{constq}) in terms of $u$ and $q$
as
\beq
u'+\frac{(1/\gth)'}{1/\gth}u=\half q_1,
\quad
q_1\equiv \vc{e}_1 \cdot \Bq.
\labl{constq1}
\eeq
Differentiating Eq.~(\ref{constq1}) once with respect to the conformal time
and using Eq.~(\ref{ueq}), we obtain the relation
\beq
\half \lh q'_1-\frac{(1/\gth)'}{1/\gth}q_1 \rh-\gksq \cH \getn q_2
=\gD u.
\labl{uqrel}
\eeq

Although the equations (\ref{constq}),(\ref{fieldeq}) and (\ref{ueq}) have
been expressed in terms of the slow-roll variables, they are exact, and no
slow-roll approximation has yet been made. Observe that, to the leading
order in slow-roll, the perturbation variables $\Bq$ and $u$ decouple, 
whereas at first order, mixing between these occur.
\subsection{Quantization of the scalar perturbations}
\labl{qtmpert}
We now study the quantization of the density perturbations described by the
Lagrangean in Eq.~(\ref{fieldaction}). We start by
introducing the matrix $Z_{mn}$ defined as
\beq
(Z)_{mn}=-(Z^T)_{mn}=\frac{1}{\cH}\vc{e}_m \cdot \cDe \vc{e}_n,
\labl{zmn}
\eeq
where the second equality follows from $\cDe(\vc{e}_m\cdot\vc{e}_n)=0$. Thus
$Z$ is antisymmetric and traceless $(\mbox{Tr} Z=0)$. Upon expanding
$\Bq=q_m \vc{e}_m$, using the basis $\{\vc{e}_m\}$, it follows from
Eq.~(\ref{fieldaction}),
\beq
\mathscr L=\half(q'+\cH Z q)^T(q'+\cH Z q)+\half q^T(\gD-\cH^2 \gO)q,
\labl{lagrange1}
\eeq
where $(\gO)_{mn}=\vc{e}^{\dag}_m \BgO\vc{e}_n$, and for notational ease, we
have suppresed the indices $m,\ n$. It will prove convenient to reduce the
Lagrangean (\ref{lagrange1}) to the canonical form. We redefine $q$ using a
new matrix $R$ as
\beq
q(\get)=R(\get)Q(\get),
\quad
R'+\cH ZR=0,
\quad
\tilde{\gO}=R^T\gO R.
\labl{qredef}
\eeq
From the equation of motion (\ref{qredef}) for $R$, it follows that $R^TR$ 
and $\mbox{det}R$ are constants, so that $R$ represents a rotation. Without 
any loss of generality, the initial value of $R$ may be chosen as 
$R(\get_0)=\openone$.
Substituting
the variables defined in Eq.~(\ref{qredef}) into Eq.~(\ref{lagrange1}) yields
\beq
\mathscr L=\half Q'^T Q'+\half Q^T(\gD-\cH^2 \tilde{\gO})Q.
\labl{lagrange2}
\eeq

To proceed with the quantization,
we employ the canonical quantization procedure to the Lagrangean
(\ref{lagrange2}). The momentum $\gP$ canonically conjugate to $Q$ is
\beq
\gP(\get,\vc{x})=\partial \mathscr L /\partial Q'^T = Q'(\get,\vc{x}).
\labl{conjmtm}
\eeq
The Hamiltonian is then given by
\beq
\mathscr H=\half \gP^T\gP-\half Q^T(\gD-\cH^2 \tilde{\gO})Q.
\labl{hamilton}
\eeq
The canonically conjugate variables $(Q,\gP)$ are promoted to quantum
operators $(\hat{Q},\hat{\gP})$ satisfying the commutation relations
\equ{
& & [\ga^T\hat{Q}(\get,\vc{x}),\gb\hat{Q}(\get,\vc{x'})]=
[\ga^T\hat{\gP}(\get,\vc{x}),\gb\hat{\gP}(\get,\vc{x'})]=0, \nonumber
\\
& & [\ga^T\hat{Q}(\get,\vc{x}),\gb\hat{\gP}(\get,\vc{x'})]=
i\ga^T\gb \gd(\vc{x}-\vc{x'}),
\labl{commute1}
}
where the delta function is normalized as
\beq
\int \gd (\vc{x}-\vc{x'})\sqrt{g^{(3)}}d^3\vc{x},
\labl{delta}
\eeq
and we have introduced the vectors $\ga,\ \gb$ with components $\ga_m,\ \gb_m$
in the basis $\{\vc{e}_m\}$ to avoid writing the indices $m,\ n$ in
the commutators. Since we are considering spatially flat hypersurfaces $(K=0)$,
the operator $\hat{Q}$ may be expanded in a plane wave basis as
\beq
\hat{Q}=\int \frac {d^3\Bk}{(2\gp)^{3/2}}
\left[Q^*_{k}(\get)\hat{a}_{\Bk}e^{i\Bk \cdot x}
+Q_{k}(\get)\hat{a}^\dag_{\Bk}e^{-i\Bk\cdot x}\right],
\labl{modexpand}
\eeq
with a similar expansion for $\hat{\gP}$. It immediately follows from
Eq.~(\ref{qredef}) that $q$ must now be interpreted as the operator
$\hat{q}$ with modes
\beq
q_{k}(\get)=R(\get)Q_{k}(\get),
\labl{qmodes}
\eeq
satisfying a mode expansion identical to Eq.~(\ref{modexpand}).
The creation and annihilitation operators $\hat{a}^\dag_{\Bk}$
and $\hat{a}_{\Bk}$ satisfy
\equ{
& & [\ga^T \hat{a}_{\Bk},\gb\hat{a}_{\Bk '}]
=[\ga^T \hat{a}^\dag_{\Bk},\gb \hat{a}^\dag_{\Bk '}]=0, \nonumber
\\
& & [\ga^T \hat{a}_{\Bk},\gb \hat{a}^\dag_{\Bk '}]=\ga^T\gb \gd(\Bk-\Bk').
\labl{commute2}
}
In order that the commutation relations (\ref{commute1}) and
(\ref{commute2}) be consistent, the following Wronskian condition
must be satisfied,
\beq
 W\{Q_{k},Q^*_{k}\} \equiv Q'_{k}(\get)Q^*_{k}(\get)
-Q'^*_{k}(\get)Q_{k}(\get)=i.
\labl{wronski}
\eeq
From the mode expansion (\ref{modexpand}) and the
Hamiltonian (\ref{hamilton}), it follows that the equation of motion for
$Q_{k}$ is
\beq
Q''_{k}+(k^2+\cH^2\tilde{\gO})Q_{k}=0.
\labl{Qeqn}
\eeq
It may be easily verified using Eq.~(\ref{Qeqn}) that the Wronskian
satisfies $dW\{Q_{k},Q^*_{k}\}/d\get=0$.

We also interpret the variable $u$ introduced in Eq.~(\ref{sasaki2}) as an
operator $\hat{u}$, and after performing a mode expansion identical to that
of $\hat{Q}$ in Eq.~(\ref{modexpand}), it follows from Eq.~(\ref{ueq}) that
the modes $u_k$ satisfy
\beq
u''_{k}+\lh k^2-\frac{\gth''}{\gth}\rh u_k
=\gksq \cH \getn q_{2k},
\quad
q_{2k}\equiv (\vc{e}_2\cdot\vc{e}_m)q_k,
\labl{ueqk}
\eeq
or, equivalently, from Eq.~(\ref{uqrel}),
\equ{
& & \gksq \cH \getn q_{2k}-\half \lh q'_{1k}
-\frac{(1/\gth)'}{1/\gth}q_{1k} \rh
=k^2 u_k, \nonumber
\\
& & q_{1k}\equiv (\vc{e}_1\cdot\vc{e}_m)q_k.
\labl{ueqk1}
}
\subsection{First order solution}
\labl{soln}
As a prelude to presenting the solution of the perturbation
equations to the first order in slow-roll, we briefly 
discuss the issue of the ambiguity in
the choice of the vacuum state when quantizing fields in an expanding
FRW background \cite{birrell}. In ordinary Minkowski space-time, there
exists a unique time direction, as well as distinct time-invariant
positive- and negative-frequency modes. However, when quantizing in a
curved background, there is neither a distinct time direction, nor a
notion of time-invariant mode. Consequently, there is no unique
vacuum state either. Hence if we have a positive-frequency mode
$Q^{+}_k$ at times $\get < \get_0 < 0$, with the initial vacuum state
$\hat{a}_{\Bk}|0\rangle=0$, then at later times
$\get > |\get_0|$, the modes will in general be described by a linear
superposition of positive- and negative-frequency modes $Q^{+}_k,\ Q^{-}_k$,
related by means of a Bogoliubov transformation:
\equ{
& & Q_{k}(\get)=\gl_{k}(\get)Q^{+}_{k}(\get_0)+
\gm_{k}(\get)Q^{-}_{k}(\get_0), \nonumber
\\
& & |\gl_{k}(\get)|^2-|\gm_{k}(\get)|^2=1.
\labl{bogo}
}
For the coefficients $\gl_k$ and $\gm_k$, one often makes the choice of the
initial values as
\beq
\gl_{k}(\get_0)=1,
\quad
\gm_{k}(\get_0)=0,
\labl{bogocoefft}
\eeq
for the adiabatic vacuum (or the Bunch-Davies vacuum in de-Sitter space), which
corresponds to the positive-frequency solution in the Minkowski space.

To proceed further, it is convenient to introduce the time $\getH$ when the
mode with wave number $k$ crosses the Hubble radius during inflation, 
so that the relation
\beq
\cH(\getH)=k
\labl{etaH}
\eeq
is satisfied for each $k$. Consequently, the inflationary epoch can be 
separated into three regions:
the \textit{sub-horizon} region ($\cH \ll k$), the \textit{transition}
region ($\cH \sim k$), and the \textit{super-horizon} region ($\cH \gg k$). We
now discuss each of these in turn.

In the sub-horizon region, we solve Eq.~(\ref{Qeqn}) with the 
$\cH^2 \tilde{\gO}$ term subdominant compared to $k^2$. The solution is obtained
in the limit $k/\cH \rightarrow \infty$ for fixed $k$ as
\beq
Q_k(\get)=\frac{1}{\sqrt{2k}}e^{ik(\get-\get_0)},
\quad
R(\get_0)=\openone.
\labl{qsubsol}
\eeq
Since one is usually interested in calculating quantities at the end of  
inflation, this region is therefore irrelevant.

We consider next the transition region. It will prove useful to introduce
the time $\get_-$ when the sub-horizon epoch ends and the transition region
begins. In a sufficiently small interval around $\getH$ we can then 
apply slow-roll to the Eq.~(\ref{Qeqn}) keeeping all the terms, but taking
the slow-roll functions to be constant to the first order. The initial
conditions are chosen as
\beq
Q_k(\get_-)=\frac{1}{\sqrt{2k}}\openone,
\quad
Q'_k(\get_-)=\frac{i\sqrt k}{\sqrt 2}\openone,
\quad
R(\get_-)=\openone.
\labl{transinit}
\eeq
Integrating the relation for $\cH'$ in Eq.~(\ref{useful}) with the initial
conditions for the transition region, we obtain
\beq
\cH(\get)=\frac{-1}{(1-\geH)\get},
\quad
\getH=\frac{-1}{(1-\geH)k},
\labl{Hexpr}
\eeq
so that $\cH(\getH)=k$.
Differentiating $\gth$ in Eq.~(\ref{ueq}) yields
$\gth '/\gth=-\cH(1+\ge+\getp)$, which can be integrated with the result
\equ{
& & \gth(z)=\gthH \lh\frac{z}{\zH}\rh^{(1+2\geH+\get^{\parallel}_\cH)},
\quad
z\equiv k\get, \nonumber
\\
& & \gthH=\frac{\gk}{\sqrt2}\frac{H_\cH}{k\sqrt \geH},
\quad
z_\cH \equiv k\getH
\labl{thetasol}
}
The differential equation (\ref{qredef}) for the rotation matrix $R$ can be
solved with the initial conditions (\ref{transinit}) leading to
\beq
R(z)=\lh\frac{z}{z_-}\rh^{-(1-\geH)^{-1}Z_\cH},
\quad
z_-\equiv k\get_-.
\labl{Rsol}
\eeq
Since the time-dependent terms in the matrix $\gO$ in Eq.~(\ref{omega}) are of
first order, we can take $\gO=\gO(\getH)\equiv \gO_\cH$ in the transition 
region. Then the matrix $\tilde{\gO}$ is given by
\equ{
\tilde{\gO}&=&R^{-1}(z)\gO_{\cH} R(z)=
\gO_{\cH}-[\gO_{\cH},Z_\cH]\mbox{ln}\frac{z}{z_-} \nonumber
\\
&=&\gO_{\cH}+3[\gd_{\cH},Z_{\cH}]\lh \mbox{ln}\frac{z}{z_{\cH}}+
\frac{3}{4}\mbox{ln}\;\geH\rh,
\labl{omegasol}
}
with
\equ{
\gd(\get)&=&-\frac{1}{3}\lh2\openone+\frac{\gO}{(1-\ge)^2}\rh \nonumber
\\
&=&
\ge \openone -\frac{a^2 M^2}{3\cH^2}+
2\ge (\vc{e}_1 \cdot \vc{e}_m)(\vc{e}_1 \cdot \vc{e}_n)^T, \nonumber
\\
\gd_\cH&=&\gd(\getH).
\labl{deltaH}
}
Here the second equality is valid to the first order in slow-roll, and we have
used the notation $M^2\equiv \vc{e}^{\dag}_m \mx M^2 \vc{e}_n$. We also made
the assumption that those components of $a^2 M^2/\cH^2$ which cannot
be expressed in terms of the slow-roll variables are of first order.
Because $\gd_\cH$ and $Z_\cH$ are both of first order, we can
take $\tilde{\gO}=\gO_\cH$ in Eq.~(\ref{omegasol}) to be a first
order quantity.

In order to write the equation for the mode $Q_k$ in the transition
region, we will find it convenient to define
$\bar{Q}_k \equiv R_\cH Q_k(z)$
and $\bar{\gO}=R_\cH \tilde{\gO}R^{-1}_\cH$, with $R_\cH\equiv R(z_\cH)$.
From Eq.~(\ref{Rsol}), we have to the first order, $Q_k(z)=\bar{Q}_k(z)$, while
from Eq.~(\ref{omegasol}) we conclude that $\bar{\gO}=\gO_\cH$ within a 
small region
around $z_\cH$. Using the above results in Eq.~(\ref{Qeqn}), the mode
equation for $Q_k$ may be written in terms of $\bar{Q}_k$ as
\beq
\bar{Q}_{k,\;zz}+\lh\openone -\frac{\gn^2_\cH-\frac{1}{4}}{z^2}\rh \bar{Q}_k=0,
\quad
\gn^2_\cH=\frac{9}{4}\openone+3\gd_\cH.
\labl{hankel}
\eeq
This equation is similar to the one obtained for the single-field inflation,
except that this is a matrix equation. 
The solution is then given in terms of
the Hankel functions of matrix valued order $\gn_\cH$,
\equ{
& & \bar{Q}_{k}(z)=\sqrt{z}[c_1(k)H^{(1)}_{\gn_\cH}(z)
+c_2(k)H^{(2)}_{\gn_\cH}(z)], \nonumber
\\
& & \gn_\cH=\frac{3}{2}\openone+\gd_\cH.
\labl{hankelsol}
}

We wish to match the solution in Eq.~(\ref{hankelsol}) so that in the limit
$k/\cH \rightarrow \infty$, the modes approach plane waves, 
$\bar{Q}_k(z)=e^{iz}/\sqrt{2k}$, see (\ref{qsubsol}). For $|z|\gg 1$, the
Hankel functions have the asymptotic forms,
\equ{
& & H^{(1)}_{\gn_\cH}(z)\sim \sqrt{2/(\gp z)}e^{i\{z-(\gn_\cH + 1/2)\gp /2\}},
\nonumber
\\
& & H^{(2)}_{\gn_\cH}(z)\sim \sqrt{2/(\gp z)}e^{-i\{z-(\gn_\cH + 1/2)\gp /2\}}.
\labl{hankelasymp}
} 
We set $c_1(k)=\sqrt{\gp/(4k)}e^{i(\gn_\cH + 1/2)\gp /2}$, and $c_2(k)=0$. 
The phase factor of $c_1(k)$ is chosen in order to match with
Eq.~(\ref{qsubsol}) at short scales, while
the factor of $\sqrt{\gp/(4k)}$ ensures conformity with the Wronskian in
Eq.~(\ref{wronski}). Therefore the final solution with the appropriate
normalization is
\beq
\bar{Q}_{k}(z)=\sqrt{\gp/(4k)} e^{i(\gn_\cH + 1/2)\gp /2}
\sqrt{z}H^{(1)}_{\gn_\cH}(z).
\labl{qbarfinal}
\eeq
It is worth mentioning that that the matrix valued Hankel functions are to be
interpreted as series expansions, just like the usual Hankel functions.

We finally discuss the solution in the super-horizon region. On super-horizon
scales we have $|z|\ll 1$, for which the asymptotic form of the Hankel 
function is
\beq
H^{(1)}_{\gn_\cH}(z)\sim \sqrt{2/\gp}e^{-i\gp/2}2^{\gn_\cH-3/2}
\frac{\gG(\gn_\cH)}{\gG(3/2)}z^{-\gn_\cH},
\labl{hankelsuper}
\eeq
so that the asymptotic solution for $\bar{Q}_k(z)$ in the super-horizon 
region is given by 
\equ{
\bar{Q}_k(z)&\sim& (1/\sqrt{2k})e^{i(\gn_\cH-1/2)\gp/2}2^{\gn_\cH-3/2}
\frac{\gG(\gn_\cH)}{\gG(3/2)}z^{\half \openone -\gn_\cH}, \nonumber
\\
&\sim&-(1/\sqrt{2k})e^{i(\gn_\cH- 1/2+2\gd_\cH)\gp /2}E_\cH
(z/z_\cH)^{-\openone - \gd_\cH }, 
\nonumber \\
\labl{qbarsuper}
}
where
\beq
E_\cH \equiv (1-\geH)\openone + (2-\ggam_E - \mbox{ln}\;2)\gd_\cH,
\labl{EH1}
\eeq
and $\ggam_E \approx 0.5772$ is the Euler constant.

In this region since $k/\cH \rightarrow 0$,
we can also solve Eq.~(\ref{ueqk}) ignoring the $k^2$ dependent term, 
leading to
\equ{
& &u_k(\get)=u_{P\;k}+C_k \gth 
+ D_k \gth \int^{\get}_{\getH}\frac{d\get'}{\gth^{2}(\get')}, \nonumber
\\
& &u_{P\;k}=\gth \int^{\get}_{\getH}\frac{d\get'}{\gth^2}
\int^{\get'}_{\getH}d\get''\gth \gksq \cH \getn q_{2k},
\labl{ueqksol}
}
where $C_k$ and $D_k$ are constants of integration, and $u_{P\;k}$ is a
particular solution. Note that since $\gth$ is a rapidly decaying function, 
we can ignore $C_k$ compared to $D_k$. In the same approximation, 
the solution of Eq.~(\ref{ueqk1}) is  
\beq
q_{1k}=d_{k}(1/\gth)+2(1/\gth)\int^{\get}_{\getH}d\get'\gth 
\gksq \cH \getn q_{2k}.
\labl{ueqk1sol}
\eeq
From Eq.~(\ref{constq1}) we see that the integration constants $D_k$ and $d_k$
are related by $D_k=\half d_k$. Considering the region where $\get$ is
sufficiently close to $\getH$, the integral in Eq.~(\ref{ueqk1sol}) may then 
be neglected, so that using Eq.~(\ref{thetasol}), we can write $q_{1k}=2D_k (1/\gth_\cH)(z/z_\cH)^{-1}$.
Taking into account the asymptotic solution (\ref{qbarsuper}), 
and the fact that $q_k=(\vc{e}_1 \cdot \vc{e}_m)^{T}q_{1k}$, we finally
obtain,
\beq
D_k=-(1/2\sqrt{2k})e^{i(\gn_\cH- 1/2+2\gd_\cH)\gp /2}
\gth_\cH (\vc{e}_1 \cdot \vc{e}_m)^{T} E_\cH.
\labl{Dksol}
\eeq
Thus the integration constant in Eq.~(\ref{ueqksol}) is
completely determined to first order in slow-roll. Inserting the result
(\ref{Dksol}) for $D_k$ in (\ref{ueqksol}), and using the relation
$a_{\cH}H_{\cH}=k$, we finally arrive at
\begin{widetext}
\beq
u_k=-\frac{1}{(2k)^{3/2}}e^{i(\gn_\cH- 1/2+2\gd_\cH)\gp /2}
\frac{H_\cH}{\sqrt{\geH}}\left[\cA(t_{\cH},t)(\vc{e}_1 
\cdot \vc{e}_m)^{T}+\cB(t_{\cH},t) \right]E_\cH,
\labl{uksolfinal}
\eeq
where we ignored $C_k$, and
\equ{
& & \cA(t_{\cH},t)=\frac{1}{a\sqrt{\ge}}\int^{t}_{t_\cH}dt'a\lh\frac{1}{H}\rh^{.},
\quad
\cB(t_{\cH},t)=\frac{1}{a\sqrt{\ge}}\int^{t}_{t_\cH}dt'a\lh\frac{1}{H}\rh^{.}
\cU(t_{\cH},t),                    \nonumber
\\
& & \cU(t_{\cH},t)=2\gksq\int^{t}_{t_\cH}dt'H\getn\sqrt{\frac{\geH}{\ge}}
\frac{a_\cH}{a}(\vc{e}_2\cdot \vc{e}_m)^{T}R\frac{Q_k}{Q_{k\cH}}.
\labl{ABU}
}
\end{widetext}
Here $Q_{k\cH}$ is the value of the asymptotic solution (\ref{qbarsuper}) for
$Q_k$ evaluated at $\get = \getH$. Observe that the solution 
(\ref{uksolfinal}) for $u_k$ is expressed entirely in terms 
of background quantities and comoving time.
This concludes our
discussion of scalar perturbations in multiple field inflation.
\subsection{Vector and tensor perturbations}
For the sake of completeness, we now present a brief discussion of
vector- and tensor-type perturbations. From the $G^0_{\; i}$ component of
Eq.~(\ref{tmunu}), together with Eq.~(\ref{tmnharmonic}), we have 
\beq
\half k^2 \gPs^{(v)}=\gksq a^2 (\gm + p)v^{(v)},
\labl{vectorpert1}
\eeq 
while the condition $T^{\gm}_{i;\gm}=0$ yields
\beq
\frac{1}{a^4}\left[a^4 (\gm+p)v^{(v)}\right]'=-\half k\gp^{(v)}.
\labl{vectorpert2}
\eeq
Equations (\ref{vectorpert1}) and (\ref{vectorpert2}) describe the 
vector-type, or rotational perturbations. Since vector sources are 
absent when the matter 
sector is composed entirely of scalar fields, the vector-type perturbations
are therefore irrelevant in the inflationary scenario.

The equation for the tensor-type, or gravitational wave
perturbations follows from the $G^i_{\;j}$
component of (\ref{tmunu}):
\beq
C^{(t)\prime \prime}+2\cH C^{(t)\prime}
+k^2 C^{(t)}=\gksq a^2 \gp^{(t)}.
\labl{tensorpert1}
\eeq 
For scalar fields we have $\gp^{(t)}=0$. We can recast Eq.~(\ref{tensorpert1})
as
\beq
v_t'' +\lh k^2-\frac{a''}{a}\rh v_t=0,
\quad
v_t=aC^{(t)},
\labl{tensorpert2}
\eeq
which is of a form similar to Eq.~(\ref{ueqk}). The solution in the
large-scale limit is then obtained by ignoring the $k^2$ 
dependent term:
\beq
C^{(t)}(\get,\vc{k})=A_k+B_k\int^{\get}\frac{1}{a^2}d\get,
\labl{tensorsol}
\eeq
where $A_k$ and $B_k$ are integration constants. Observe that $\gPs^{(v)}$,
$v^{(v)}$, $\gp^{(v)}$ and $C^{(t)}$ appearing in these equations are
gauge-invariant, see Eq.~(\ref{transforms}).
It is interesting to note that we can derive the equations for vector and
tensor perturbations without taking into account the scalar fields. 
Therefore the presence of scalar fields do not formally affect these
perturbations.

\section{Conclusion}
\labl{discuss}
In this paper we presented a general framework for analyzing linear
cosmological perturbations in the multicomponent inflation scenario
using the gauge-ready approach. Our model consists of multiple
scalar fields induced with a positive-definite, but a general
field metric, coupled non-minimally to Einstein gravity. The space-time
metric is chosen as the perturbed FRW world model. We gave the complete
set of perturbation equations in the gauge-ready form, and derived a
set of equations for gauge-invariant perturbed order variables. We wrote
the equations governing scalar perturbations using generalized forms of
slow-roll variables. We then applied canonical quantization to the
scalar perturbations and obtained the solutions to the first order in
slow-roll. We found that the asymptotic solutions in the super-horizon
region are given in terms of Hankel functions. This is similar to the
single-field inflation case, except that the order of the Hankel functions
is matrix valued.

There are a number of possible extensions to our work. First, the immediate
next step would be to calculate the power-spectra and spectral indices
in realistic models of multiple-field inflation. Second, it would be
interesting to interface our approach with the CMBFAST \cite{cmbfast} or
the CAMB \cite{camb} computer codes and compare with the WMAP results.
Third, the methods in this paper may be extended to include generalized
gravity as well as hydrodynamical fluids. Fourth, a systematic investigation
of adiabatic and isocurvature perturbations in the context of multicomponent
inflation can be made using the gauge-ready approach. These, and other issues,
will be presented in a future work.

\end{document}